\documentclass[12pt,fleqn]{article}

\usepackage{amsmath}
\usepackage{amsfonts}
\usepackage{graphicx}
\usepackage{dcolumn}
\usepackage{upgreek}
\usepackage{latexsym,slashed}
\usepackage{amssymb,amsfonts}
\usepackage{cancel}
\usepackage{bm}        
\usepackage{amsmath,mathrsfs}   

\newcommand{\pacs}[1]{\smallskip\noindent{\sl PACS numbers:
                       \hspace{0.3cm}#1}\par\bigskip\rm}

\newcommand{\address}[1]{\begin{center}\large #1\end{center}}

\def\beq{\begin{eqnarray}}
\def\eeq{\end{eqnarray}}

\def\p{\partial}

\def\R{{\hbox{{\rm I}\kern-.2em\hbox{\rm R}}}}   
\def\H{{\hbox{{\rm I}\kern-.2em\hbox{\rm H}}}}   
\def\N{{\hbox{{\rm I}\kern-.2em\hbox{\rm N}}}}   
\def\C{{\ \hbox{{\rm I}\kern-.6em\hbox{\bf C}}}} 
\def\Z{{\hbox{{\rm Z}\kern-.4em\hbox{\rm Z}}}}   

\catcode`\@=11
\@addtoreset{equation}{section}

\begin{document}
\tolerance=5000

\title{Hamilton-Jacobi Tunneling Method for Dynamical Horizons in
  Different Coordinate Gauges} 

\author{Roberto~Di~Criscienzo$\,^{(a),
  (b)}$\footnote{rdicris@science.unitn.it}, 
  Sean~A.~Hayward$\,^{(b)}$\footnote{sean\underbar{ }a\underbar{
  }hayward@yahoo.co.uk},\\ 
Mario~Nadalini$\,^{(a)}$\footnote{nadalini@science.unitn.it},
  Luciano~Vanzo$\,^{(a)}$\footnote{vanzo@science.unitn.it} and
  Sergio~Zerbini$\,^{(a)}$\footnote{zerbini@science.unitn.it}}
\date{}
\maketitle
\address{$^{(a)}$ Dipartimento di Fisica, Universit\`a di Trento \\
and Istituto Nazionale di Fisica Nucleare - Gruppo Collegato di Trento\\
Via Sommarive 14, 38100 Povo, Italia}
\address{$^{(b)}$ Center for Astrophysics - Shanghai Normal University,\\
100 Guilin Road, Shanghai 200234, China}
\medskip \medskip

\begin{abstract}
Previous work on dynamical black hole instability is further
elucidated within the Hamilton-Jacobi method for horizon tunneling
and the reconstruction of the classical action by means of the
null-expansion method. Everything is based on two natural
requirements, namely that the tunneling rate is an observable and
therefore it must be based on invariantly defined quantities, and
that coordinate systems which do not cover the horizon should not be
admitted. These simple observations can help to clarify some
ambiguities, like the doubling of the temperature occurring in the
static case when using singular coordinates and the role, if any, of
the temporal contribution of the action to the emission rate. The
formalism is also applied to FRW cosmological models, where it is
observed that it predicts the positivity of the temperature
naturally, without further assumptions on the sign of energy.
\end{abstract}


\section{Introduction}

In  previous papers \cite{DiCriscienzo:2007fm,Hayward:2008jq} we
considered the quantum instability of dynamical black hole using a
variant of the tunneling method introduced by Parikh and Wilczek
(PW) in the static case \cite{parikh,P} to uncover aspects of the
back reaction effects; see also \cite{obr}. The PW tunneling method
was refined and extended to more general cases in \cite{svmp} and
other papers \cite{V,others}, including  its relation with
thermodynamics \cite{Pilling:2007cn}. Other  points of view and
criticisms can be found in \cite{anm}.

In the present paper, namely in section \ref{tunneling}, we review
this variant method, which in the static case goes under the name of
Hamilton-Jacobi (HJ) method of tunneling, and was introduced in
\cite{Visser:2001kq,angh:2005}. A complete comparison analysis with
the Parikh-Wilczek method has been presented in \cite{mann}. The HJ
tunneling (across a horizon) connects the invariant surface gravity
to a dynamical local temperature through the leading term in the
black hole tunneling rate (\ref{prob}).

First, we would like to
notice that the main novelty of the HJ method is its manifest
covariance, compared to the original Parikh-Wilzcek approach
\cite{parikh,P} where a non manifestly covariant Hamiltonian
formulation was used. With regard to this issue, it is our opinion
that HJ method is particularly suitable for the generalization to
the dynamical spherically symmetric case, as we will try to
substantiate in this paper.

The tunneling method provides not only new physical insight to an
understanding of the black hole radiation, but is also a powerful
and simple way to arrive at an expression for the surface gravity
for a vast range of solutions, especially non-stationary and, in our
case, dynamical black holes. This looks important since several
definitions of the surface gravity for evolving horizons have been
proposed in the past, all fitting some kind of first law of black
hole mechanics more or less equally well. Still different results
are advocated for expanding cosmological black holes in
\cite{Faraoni:2007gq,Saida:2007ru}. A comparison is discussed
thoroughly in \cite{Nielsen:2007ac,Vanzo:2008uq}, but we anticipate
that Hayward's definition \cite{Hayward:1997jp} of surface gravity,
related to Kodama's theory \cite{Kodama:1979vn} of spherically
symmetric space-times, is the most interesting to us since it is the
one to which the tunneling method naturally leads. \\

As we said above, a way to understand Hawking radiation is by means
of tunneling of particles through black-hole horizons. Such a
tunneling approach uses the fact that the WKB approximation of the
tunneling probability for the classically forbidden trajectory from
inside to outside the horizon is:
\begin{equation}
\Gamma \propto \exp\left(- \frac{2}{\hbar}\mbox{Im } I\right),
\label{prob}
\end{equation}
where $I$ is the classical action of the (massless) particle, to
leading order in $\hbar$. It is of the utmost importance that the
exponent be a scalar invariant, otherwise no physical meaning can be
given to $\Gamma$. If, in particular, it has the form of a thermal
emission spectrum with $2 \mbox{Im }\, I=\beta \omega $, then the
inverse temperature $\beta$ and the particle's energy $\omega$ have
to be separately scalars, since otherwise no invariant meaning can
be given to the horizon temperature, which would not be an
observable. If, even in the presence of reasonable physical
conditions, more than one prescription for defining an invariant
energy is available, then also more notions of invariant temperature
will exist and further analysis, or observations, could be needed.
In most cases this could only be a scale transformation or a choice of
a different family of observers.\\ 
In principle, all the Standard Model particles are expected in the
Hawking radiation spectrum. However, most of the calculations in the
literature have been performed just for scalar fields. Spin one-half
emission was considered in \cite{Kerner:2007rr} for stationary black
holes, in \cite{Li:2008ws} for the special case of the BTZ black
hole and finally \cite{DiCriscienzo:2008dm} studied the case of
evolving horizons. \\
In Ref.\ \cite{Hayward:2008jq} we have already discussed the
Hamilton-Jacobi tunneling method for arbitrary spherically
symmetric dynamical black holes. In this paper, we would like to
present a systematic derivation of the results, pointing out the key
points and making use of a covariant coordinate
approach\footnote{For static black holes
  a coordinate-free formulation has been given by Stotyn and
  co-workers \cite{Stotyn:2008qu}.} which is particularly convenient
in the discussion of several explicit examples we will consider as
applications of the general method. We point out that our
description of the tunneling rate is not equivalent to the
``canonically invariant'' description due to Chowdhury
\cite{Chowdhury:2006sk}, which in fact obtains half the correct
temperature for black holes and infalling shells. Probably a
comparison analysis would be worthy, at this point. \vspace{0.3cm}
\\
Recall that any spherically symmetric metric can locally  be
expressed in the  form
\beq
\label{metric}
ds^2 =\gamma_{ij}(x^i)dx^idx^j+ R^2(x^i) d\Omega^2\,,\qquad i,j \in \{0,1\}\;,
\eeq
where the two-dimensional metric
\beq d\gamma^2=\gamma_{ij}(x^i)dx^idx^j
\label{nm}
\eeq
is referred to
as the normal metric, $x^i$ are associated coordinates and $R(x^i)$ is the
areal radius, considered as a scalar field in the normal two-dimensional space.
Another relevant scalar quantity on this normal space is \beq
\chi(x)=\gamma^{ij}(x)\partial_i R\partial_j R\,. \label{sh} \eeq The dynamical
trapping horizon, say $H$, may be defined by
\beq \chi(x)\Big\vert_H = 0\,, \qquad \partial_i\chi \Big\vert_H
\neq 0\,. \label{ho} \eeq
This is equivalent to the vanishing of the expansion $\theta_{(\ell)}$
of the null, future directed congruence which is normal to a section of the
horizon. In general there will be another null congruence of
``incoming rays'' with a related expansion $\theta_{(n)}$: if
$\theta_{(n)}<0$  and  also ${\cal L}_n\theta_{(\ell)}<0$ along $H$, then
the horizon is of the future, outer type. Most of our results will be
valid for this type of trapping horizon.

The Misner-Sharp gravitational mass, in units $G=1$, is defined by
\beq
m(x)=\frac{1}{2} R(x)\left(1-\chi(x) \right)\,. \label{MS}
\eeq
 This is an invariant quantity on the normal space. Note also that, on
 the horizon, 
$m\vert_H=m=R_H/2$. Furthermore, one can introduce a dynamic surface
gravity \cite{Hayward:2008jq} associated with this dynamical
horizon, given by the normal-space scalar \beq
\kappa_H=\frac{1}{2}\Box_{\gamma} R
\Big\vert_H=\frac{1}{2\sqrt{-\gamma}}\partial_i(\sqrt{-\gamma}
\gamma^{ij}\partial_jR)\Big\vert_H\,. \label{H} \eeq Recall that, in
the spherical symmetric dynamical case, it is possible to introduce
the Kodama vector field $K$, such that $(K^\alpha G_{\alpha
\beta})^{;\beta} =0$ that can be taken as its defining property.
Given the metric (\ref{metric}), the Kodama vector components are
\beq K^i(x)=\frac{1}{ \sqrt{-\gamma}}\varepsilon^{ij}\partial_j R\,,
\qquad K^\theta=0=K^\varphi \label{ko} \;. \eeq The Kodama vector
gives a preferred flow of time and in this sense it generalizes the
flow of time given by the static Killing vector in the static case.
As a consequence, we may introduce the invariant energy associated
with a particle  by means of the scalar quantity on the normal space
\beq \label{e} \omega =- K^{i}\partial_i I\,, \eeq where $I$ is the
classical action of the massless particle, which we assume to
satisfy the Hamilton-Jacobi equation \beq \label{hj}
\gamma^{ij}\partial_i I \partial_j I=0\,. \eeq We stressed above the
importance to have at disposal an invariant definition of energy.
Eq.~\eqref{e} certainly satisfies this requirement if the action is,
as it normally is, a scalar. In the following our aim will be not so
much a detailed picture of the physical process of horizon
tunneling, as to show that there is a precise invariant
prescription to deal with the imaginary part of the action, in case
there is one, which is valid for all solutions in all coordinates
systems which are regular across the horizon. \vspace{0.3cm}
\\
The plan of the paper is the following. In Section II we discuss the
role of coordinate invariance in the Hamilton-Jacobi method.
 In Section III, the first law for dynamical black holes is derived.
 In Section IV, FRW space-times are
 investigated in detail as relevant examples of dynamical horizons.
 The naturally obtained positivity of the temperature is emphasized. In
 Section V, static black holes are revisited,
 and the issue of Hayward surface gravity and Killing surface gravity is
 discussed in detail. The paper ends with the Conclusions.

\section{Hamilton-Jacobi Tunneling and Gauge
Invariance}\label{tunneling}

To begin with, we would like to point out that, within this context,
the tunneling 
has to be interpreted in a different way from the usual textbook
treatment of tunneling effect in (non-relativistic) quantum
mechanics. In fact, strictly speaking, there is no barrier here: an
issue which has been clarified by Parikh and dubbed by him as
``secret tunneling'' \cite{P}. Furthermore, the appearance of an
imaginary part in the ``classical'' action is  due to the presence
of a dynamical horizon and the use of Feynman's
$i\epsilon$--prescription in dealing with an otherwise divergent
integral within the classical realm. This is the way, together with
equation (\ref{prob}), in which quantum effects enter into the game.

To be more specific, our key assumption is the following: that we can
reconstruct the whole action $I$, from $\partial_i I$ by means of
\beq
\label{I}
I = \int_{ \gamma}dx^{i}\partial_i I\,,
\eeq
where $\gamma$ is an oriented, {\it null curve},
with at least one point on the dynamical horizon. We can split the integration
along this null curve in two pieces, one very near to the dynamical horizon,
the remaining contribution living in the regular part of space-time. Then, we
may perform a near-horizon approximation in the first integral. As a
consequence, we have to use a system of coordinates which is regular
on the horizon, otherwise this procedure {\it cannot} be applied, due
to highly singular quantities 
involved.  This procedure permits to evaluate the integration along
what we can call, certainly with some abuse  
of notation, the \textquotedblleft radial'' and \textquotedblleft
temporal'' parts and see if an  imaginary part shows up, and what is their
relation.

In Ref.\ \cite{Hayward:2008jq}, we have presented a derivation of
the relevant imaginary  part of the classical action and tunneling
rate (\ref{prob}), valid for a future trapped horizon, which reads
\beq \mbox{Im }\, I = \mbox{Im }\, 
\left( \int_{ \gamma} dx^{i} \partial_i I
\right)=\frac{\pi\omega_H}{\kappa_H}\,, \label{ima} \eeq where
$\kappa_H$ is the dynamical surface gravity (\ref{H}), and
$\omega_H$ the Kodama energy evaluated on the dynamical horizon.
These quantities are scalars in the normal space, thus the leading
term of the
tunneling rate is invariant, as we expect to be from  an observable. \\
For the sake of completeness and clarity, we are briefly reporting the
essential steps of the derivation in several coordinate systems.
\medskip
\paragraph{The EFB Gauge.} Let us start with giving the computation as
in \cite{Hayward:2008jq}. The key point  
is the observation that it is always
possible to rewrite, locally, any spherically symmetric metric in an
Eddington-Finkelstein-Bardeen (EFB) form which is regular on a trapping
horizon. For black holes, one is concerned with future trapping horizons and so
the advanced (rather than retarded) form is used: \beq
ds^2=-e^{2\Psi}Cdv^2+2e^{\Psi}dvdr+r^2d\Omega^2\,, \eeq where we are using
$x^i=(v,r)$ as coordinates, and  $C=C(v,r)$, $\Psi=\Psi(v,r)$.   In this gauge,
life is easy since the areal radius of the spheres of symmetry $R$
turns out to be equal to the $r$-coordinate,  
i.e. $R=r$, and $\chi=C$: the horizon location is defined by the
condition $C(v,r)\vert_H=0$. 
The Kodama vector and the invariant energy assume the simple expressions given
by $K=(e^{-\Psi}, \overrightarrow 0)$, and
$\omega=-e^{-\Psi}\partial_v I$ while the invariant surface gravity is
just given by $\kappa_H=\partial_rC_H/2$. One may note that $\Psi$
transforms as an ordinary Liouville field,
i.e. $\Psi\to\Psi+\ln|\partial\tilde{v}/\partial v|$, under
$v\to\tilde{v}(v)$, making $\omega$ invariant under reparametrizations
of the advanced time coordinate.\\
Expanding along a null direction from a point on $H$ outward in a
neighbourhood of the horizon gives (since $C_H=0$)
\beq 0=e^{\Psi_H}\Delta r \Delta v\,,
 \label{ne1}
\eeq
which shows that the temporal ($v$-)contribution does not play any
role in what concerns us at the moment (the evaluation of
Im$\;I$). \\ 
From the HJ equation, we see that, for outgoing modes, \beq C
(\partial_rI)=  2\omega \;. \label{c1} \eeq Thus, taking
(\ref{ne1}) and (\ref{c1}) into account, one has
\begin{eqnarray}
\mbox{Im}\,I&=&\mbox{Im}\int_{\gamma}\left(\partial_r I dr +
\partial_v Idv\right) \nonumber \\ 
&=&  \mbox{Im} \int_{\gamma} dr\,\frac{2\omega}{C}  \label{figa} \\
&=& 2 \,\, \mbox{Im}\int_{\gamma} dr \,\frac{\omega }{\partial_r
  C\big\vert_H(r-r_H-i0)} \label{culo} \\ 
&=& \frac{\pi \omega_H}{\kappa_H}\,,
\end{eqnarray}
where again, the quantity $C$ has been expanded around the horizon along the
null direction, that is
\beq
 C(v,r)\approx \p_r C\Big\vert_H \Delta r  + \dots
\eeq
and Feynman's $i\epsilon$-prescription has been implemented in order to
deal with the simple pole (\ref{culo}). $\kappa_H= \p_r
C\big\vert_H/2$ coincides 
with our geometrical expectations and we see that, in EFB coordinate
system, the temporal integration does not give any contribution to the
imaginary part of the 
action of particles tunneling through the (trapping) horizon. \vspace{0.3cm}\\
However, for practical reasons, it might be convenient to work in
other (regular on the horizon) coordinate systems. So, denoting the
temporal and spatial coordinates by $x^i=(t,r)$, we are going to
discuss three more instructive gauges. 
\medskip
\paragraph{The r-Gauge.} The normal metric here is non-diagonal, but
as in EFB gauge, 
$R=r$. We have \beq ds^2=d\gamma^2+r^2d\Omega^2\,, \label{tnd} \eeq  where the
reduced normal metric is \beq d\gamma^2=-E(r,t)dt^2 +2 F(r,t)dtdr +
G(r,t)dr^2\,, \qquad F \neq 0\,. \label{nd} \eeq  The horizon is located where
\beq
\chi(t,r)=\gamma^{ij}\p_iR \p_j R = \gamma^{rr}(t,r)=\frac{E}{EG + F^2}
\label{l}
\eeq
 vanishes, i.e.\ at  $E_H=0$, provided
$F_H\neq 0$. The Kodama vector reads \beq K=\left(\frac{1}{\sqrt{F^2 +
EG}},\overrightarrow 0\right)\,, \eeq and the invariant energy
\beq
\omega=-\frac{\partial_t I}{\sqrt{F^2 + EG}}\,.
\label{ro}
\eeq
The dynamical surface gravity is, from \eqref{H},
\beq
\kappa_H=  \left[\frac{1}{2 F^3}\left( E' F  -\frac{1}{2} \dot E
  G\right)\right]_H \,, 
\eeq
where an overdot and a prime denote differentiation with respect to
$t$ and $r$, respectively. From the metric the null, radial, expansion
gives $\Delta t=-\frac{G}{2 F}\Big\vert_H \Delta r$ (the other
solution being related to the ingoing null ray), so 
 within the  near-horizon approximation and after some calculation we
 get $\chi\simeq 2\kappa_H(r-r_H)$; also, $\partial_t I=-F_H \omega$
 from definition \eqref{ro} and the horizon condition $E=0$.\\ 
 Splitting the integration along $\gamma$ according to what we said
 above, we end with $I$ given by the sum of a real term and a possibly
 imaginary part coming from near horizon approximation: 
\begin{eqnarray}
 I \int_{\gamma} (dr \p_r I + dt \p_t I) = \int_{\gamma} 
dr \,\left[\partial_rI +\frac{1}{2} G_H \omega_H \right]\;.
\end{eqnarray}
What is remarkable is that in this gauge, the temporal, $t$--part is
present\footnote{Since it contributes to the total action through
the $\frac{1}{2}(G \omega)_H$ term.}, but being regular, it does
not contribute to the imaginary part of the action.  Making use of
the HJ equation, the Kodama energy expression (\ref{ro}) and
equation (\ref{l}) as well,  one has \beq \chi (\partial_r
I)^2-2\frac{\omega F}{\sqrt{EG+F^2}}\partial_r I-\omega^2 G=0\,,
\eeq thus, for outgoing modes, \beq
\partial_r I=\frac{\omega F}{\sqrt{EG+F^2}\,\,
  \chi}\left(2+O(\chi)\right)\,. 
\eeq
Making use of this equation and Feynman's prescription, $E_H=0$ and
$\chi\simeq 2\kappa_H (r-r_H)$, 
one has for the outgoing mode
\beq \mbox{Im }\, I & & =\mbox{Im}\,
\int_{\gamma} dr \,\partial_r I  \nonumber \\
& & = \mbox{Im} \int_{\gamma} dr\, \frac{\omega
  F}{\sqrt{F^2 +
EG}}\frac{1+
  \sqrt{1+O(\chi)}}{2\,\kappa_H} \frac{1}{(r-r_H-i0) } \nonumber \\
& & =
\frac{\pi \omega_H}{\kappa_H}\,,
\eeq
in agreement with the EFB gauge.
\medskip
\paragraph{The Synchronous Gauge.} The second coordinate system we
would like to consider is described by the line element
\beq
ds^2=-dt^2+\frac{1}{B(r,t)}dr^2+R^2(r,t)d\Omega^2=d\gamma^2+R^2(r,t)
d\Omega^2\,, 
\eeq
in which the metric is diagonal, but the proper radius of the spheres
of symmetry $R$ is a function of the coordinates $r$ and $t$. In this
case, one has 
\beq
\chi=-(\partial_tR)^2+B (\partial_r R)^2\,,
\eeq
thus the future sheet of the trapping horizon  $\chi_H=0$ is given by
 \beq
(\partial_t R)_H=-\sqrt{B_H} (\partial_r R)_H\,, \eeq in which we
 are assuming again a 
regular coordinate system on the horizon, namely that  $B_H$ and its partial
derivatives are non-vanishing. The Kodama vector reads  \beq K=(\sqrt{B}
\partial_r R, -\sqrt{B} \partial_t R, 0 , 0)\,, \eeq and the invariant energy
\beq \omega=\sqrt{B} (-\partial_r R \partial_t I+\partial_t R \partial_r I)\,.
\eeq The dynamical surface gravity may be evaluated and reads \beq
\kappa_H=\frac{1}{4}\left(- 2\partial^2_t R_H+2 B_H \partial^2_r R_H+
\frac{1}{B_H}\partial_t R_H  \partial_t B_H +\partial_r R_H
\partial_r B_H \right)\,.
\label{sd}
\eeq
Making use of the horizon condition, we may rewrite
\beq
\kappa_H=\frac{1}{4}\left(- 2\partial^2_t R_H+2 B_H \partial^2_r R_H -
\frac{1}{\sqrt{B_H}}\partial_r R_H  \partial_t B_H +\partial_r R_H
\partial_r B_H \right)\,.
\eeq From the metric, the HJ equation reads simply \beq
-(\partial_tI)^2+B (\partial_r I)^2=0\,. \eeq As a consequence, the
outgoing temporal contribution is equal to the radial one and we
have \beq I=2\int_{\gamma} dr \partial_r I\,. \eeq The HJ equation
and the expression for the invariant energy lead to \beq
\partial_r I =\frac{\omega} {  B \partial_rR+ \sqrt{B}\partial_t R}\,.
\eeq
which has a pole on $H$. Making the expansion along the outgoing null
curve, for which $\Delta t=- \frac{1}{\sqrt{B_H}}\Delta r$,
in the near-horizon approximation, one gets
\beq
\mbox{Im }\, I=2 \cdot \mbox{Im }\, \int_{\gamma} dr \frac{\omega}{2
  \kappa_H (r-r_H-i0)} = \frac{\pi
  \omega_H}{\kappa_H}\,,
\eeq
which again coincides with the previous result. But notice that in
this gauge, the temporal contribution is essential
indeed to provide the correct result: without it the temperature would
be doubled.
\medskip
\paragraph{Conformal 2D  Gauge.} Another coordinate system where the
temporal contribution to the action plays an essential role
is the general diagonal form of a spherically symmetric metric, which
reads
\beq
ds^2=e^{\psi(t,r)}\left( -dt^2+ dr^2 \right)+R^2(t, r) d\Omega^2\,.
\label{k}
\eeq
In this form, the normal metric is conformally related to the
2-dimensional Minkowski space-time. The $\chi$ function simply reads
\beq
\chi=e^{-\psi}\left( (-\partial_{t} R)^2 +(\partial_r R)^2\right)\,,
\eeq
which leads to the (future trapped) horizon condition
\beq
(\partial_{t} R)_H = -(\partial_r R)_H\,. 
\label{hk}
\eeq
The Kodama vector and associated invariant energy are
\beq
K=e^{-\psi}\left( \partial_r R , -\partial_{t} R , 0, 0 \right)\,,
\eeq
\beq
\omega=e^{-\psi}\left(- \partial_r R \partial_{t} I + \partial_{t} R
\partial_r I \right)\,.
\label{k2} \eeq The dynamical surface gravity reads \beq
\kappa_H=\frac{1}{2} e^{-\psi_H}\left( -\partial_{t}^2 R +
\partial_r^2 R \right)\Big\vert_H\,. \label{kt} \eeq Due to
conformal invariance, the HJ equation is the same as in
two-dimensional Minkoswki space-time and we may take \beq
\partial_+I=\partial_{t} I + \partial_r I =0\,.
\label{hjk}
\eeq
Since the null expansion condition leads to $\Delta x^+=\Delta
{t}+\Delta r =0$, for outgoing modes we get 
\beq
I=\int_{\gamma} \left(dr \partial_r I+ d t \partial_{t} I
\right) = 2 \int_{\gamma} dr
\partial_r I \,.
\label{b}
\eeq
Furthermore, due to (\ref{k2}) and (\ref{hjk}), one has
\beq
\partial_r I=\frac{\omega}{e^{-\psi} (\partial_r R+\partial_{t} R)}\,.
\eeq
and we have a pole at the horizon. Making use of near horizon
approximation along the null direction, 
from  (\ref{hk}) and (\ref{kt}), one has 
$ (\partial_r R)_H+(\partial_{t} R)_H=0$, $\Delta{t}+\Delta r =0$, thus
\begin{eqnarray}
\partial_r R+\partial_{t} R &=&\left(\partial_{rr}^{2} R
-\partial_{rt}^{2} R- \partial_{tt}^{2} R +\partial_{tr}^{2} R
\right)\Big\vert_H (r-r_H) + \dots \nonumber \\ 
&=& 2 \kappa_H (r-r_H) +\dots
\end{eqnarray}
As a result, making use of Feynman's prescription, one again arrives
at  equation (\ref{ima}). \\
In the previous computations various choices of signs have been
applied in such a way that it may seem they  were chosen somewhat ad hoc in
order to get the wanted result. This is not so. Once the future
sheet of the trapping horizon has been chosen, and the sign of the
Kodama vector so determined that it is future directed, no other sign
uncertainties will occur for either outgoing or ingoing particles. On
the other hand, if there exist a past sheet in the trapping horizon
then using the tunneling picture we may as well compute the action
along an inward directed\footnote{The ambiguity inherent in this and
  analogous terms is easily resolved if the manifold is asymptotically flat.}
curve at the horizon. Then there will be 
again a non-vanishing imaginary part, but we can interpret it as a small
absorption probability.

\section{ The First Law for Dynamical Black Holes}

 Here we present, for the  sake of completeness, a derivation of a
 version of the first law. To this aim, 
 let us introduce another invariant in the
 normal space, related to the stress-energy tensor:
 \beq
 T^{(2)}=\gamma^{ij}T_{ij}\,.
 \eeq
 First, let us prove the following invariant relation
 \beq
 \kappa_H=\frac{1}{2R_H}+8\pi R_H T^{(2)}_H\,,
 \label{vanzo}
 \eeq
 valid on the dynamical horizon. \\
 We may use the \textit{EFB gauge}, in which $R=r$ and $\chi=C$ and, in this
 gauge, we have  \beq \kappa_H= \frac{1}{2} \partial_r C_H. \eeq On the other
 hand, the Einstein equations are very simple in this gauge and we have \beq
 \frac{1-C}{2}-\frac{r}{2}\partial_r C=-4 \pi r^2 T_v^v\,. \eeq Thus, on the
 horizon we get \beq \frac{1}{2}-\frac{r_H}{2}\partial_r C_H = -4 \pi r^2_H
 T_H^{(2)} \eeq
 since $T^r_{r\; H}=0$, which leads immediately to (\ref{vanzo}). \\
 However, it is easy to prove the same relation in another coordinate system,
 for example the \textit{r-gauge}. To this purpose, let us introduce
 the horizon
 area and the areal volume associated with the horizon, with their respective
 differentials:  \beq \mathcal A_H = 4\pi R_H^2\,,\qquad d \mathcal
 A_H=8\pi R_H
 dR_H\,, \eeq \beq V_H=\frac{4}{3}\pi R_H^3\,,\qquad d V_H=4\pi R_H^2 dR_H\,.
 \eeq Then a direct calculation gives \beq \frac{\kappa}{8 \pi}d \mathcal A_H
 =d\left(\frac{R_H}{2}\right) + T_H^{(2)} dV_H\,. \eeq In turn, this equation
 can be recast in the form of a first law, once we introduce the MS energy at
 the horizon:   \beq dm=\frac{\kappa}{2 \pi} d\left( \frac{\mathcal
 A_H}{4}\right) -T_H^{(2)} dV_H\,, \eeq where $S_H=\mathcal A_H/4$ generalizes
 the Bekenstein-Hawking black hole entropy .

\section{An Explicit Example: The FRW Spacetime}

A very interesting example of the tunneling method is provided by a
generic FRW space-time with flat spatial sections, \beq
ds^2=-dt^2+a(t)^2dr^2+ a(t)^2r^2 d \Omega^2\,. \label{frw flat} \eeq
At first glance, this example can seem to lie somehow outside the main
stream of the paper which, up to this point,  
has been devoted to the study of dynamical black hole
horizons. However, what we are considering is a truly dynamical
horizon but, this time, of cosmological interest.\\ 
As explained above, metric (\ref{frw flat}) belongs to the class of
Lema\^{\i}tre-Rylov, a subset of the \textit{synchronous dynamical
space-times}, and we have just showed this to be a coordinate system
where the evaluation of the imaginary part of the action gets a
contribution due to the integration along the $t$--coordinate.\\
The normal reduced metric is diagonal with coefficients $\gamma_{ij}
=$ diag$(-1, a(t)^2)$ and $\chi=1-r^2\dot{a}(t)^2$. The dynamical
horizon is the Hubble horizon (we assume $H(t)>0$, see also
\cite{cai08}) \beq r_H= 
\frac{1}{a(t) H(t)}\, \label{inner_horizon}
\eeq
with the Hubble parameter $H(t)=\dot a/a$. The dynamical surface
gravity is
\beq
\kappa_H=-\left( H+\frac{\dot H }{2 H} \right)\,,\label{frw_dyn_sg}
\eeq
and the minus sign refers to the fact the horizon in question
(\ref{inner_horizon}) is, in Hayward's terminology, of the inner
type. For example, in the Einstein-de Sitter model, with $a(t)\propto
t^{2/3}$, we would obtain
\[
\kappa_H=-\frac{1}{6t}
\]
while the de Sitter model in inflationary coordinate has $\dot{H}=0$
and $\kappa_H=-\sqrt{\Lambda/3}$, though this result is valid in any
other patch; furthermore, $\kappa_H=0$ is only
possible in a radiation dominated universe, where
$a(t)\propto\sqrt{t}$. One easily sees that in general, for a flat
model with $a(t)\propto t^n$ one has
\[
\kappa_H=-\left(n-\frac{1}{2}\right)t^{-1}
\]
so only for $n<1/2$ is our surface gravity positive. This regime
should not be physically allowed, however, since radiation dominated
models occur either with massless particles or with
ultra-relativistic massive ones, both of which are limiting cases.
It seems as if in these cases we should define the temperature as
$T= |\kappa_H|/2\pi$.  However, we will see that the tunneling
method just gives the right signs without invoking absolute values.
\vspace{0.3cm}\\
For a massless particle, the reduced HJ equation is 
\beq
 \p_t I = \frac{1}{a(t)} \p_r I\,.
\label{bb}
\eeq
The full classical
action of outgoing particles is \beq I =
\int_{ \gamma} dx^i \, \p_i I \,, \eeq with
$ \gamma$ an oriented curve with positive orientation
along the increasing values of $x^i=(t,r)$. Radially moving massless
particles follow, of course, a null direction. Then, we can perform
a null-horizon radial expansion, \beq 0 = ds^2 = -dt^2 + a^2(t) dr^2
\eeq which gives \beq \Delta t = a(t) \Delta r
\,.\label{ne} \eeq for out  particles. The
outgoing particle action, that is the action for particles coming
out of the horizon, is then
\begin{eqnarray}
I_+ &=& \int dt\, \p_t I + \int dr\, \p_r I  \label{minus}\\
&=&  2 \int dr\, \p_r I \,\label{+action}
\end{eqnarray}
upon using equation (\ref{ne}). 
The Kodama vector reads $K=(1,-rH, 0 , 0)$ and the invariant energy of
a  particle is given by 
\beq
\omega=-\partial_t I+rH \partial_r I\,.
\label{b2}
\eeq 
Thus
\beq
\p_r I=\frac{a(t)\omega}{ra(t)H-1}\,,
\eeq
and we have that \beq I_+ = -2 \int dr
\frac{a(t) \omega}{1-r a(t) H(t)}. \eeq Expanding the function
$f(r,t):= 1-r a(t) H(t)$ close to the horizon, again along a null
direction, we get
\begin{eqnarray}
f(r,t) &\approx & - a H \vert_H \Delta r_H -r \ddot a \vert_H \Delta
t_H +\dots \nonumber \\ 
&\approx& 2 a(t) \left[ - \left(H + \frac{\dot H}{2H}\right)\right]
(r-r_H) + \dots \nonumber \\ 
&\equiv& 2 \kappa_H a(t) (r-r_H) +\dots \;,\label{hay sg}
\end{eqnarray}
where $\kappa_H$ represents the (dynamical) surface gravity associated
to the horizon. \\ 
The action of the outgoing particle now reads, \beq I_+ =
 -  \int_{\gamma} dr \frac{\omega}{\kappa_H (r-r_H - i
0)}.\label{step} \eeq In order to deal with the simple pole in the
integrand, we implement the Feynman's $i\epsilon$~{-}~prescription.
In the final result, beside a real (irrelevant) contribution, we
obtain the following imaginary part: \beq \mbox{Im}\, I_+ =
-\frac{\pi \omega_H}{\kappa_H}\, .\label{im action 1} \eeq

As a consequence, we may  interpret  $T=-\kappa_H/2 \pi >0$ as
the dynamical temperature associated with FRW flat space-times. In
particular, this gives naturally a positive temperature for de
Sitter spacetime, a long debated question years ago, usually
resolved by changing the sign of the horizon's energy.  It should be
noted that in the literature, the dynamical temperature is usually
given in the form  $T=H/2\pi$, with a missing term depending on
$\dot H$,  exceptions being the papers \cite{Wu:2008rb}. Again,
$T$ becomes negative only for the unphysical flat models with
$n<1/2$, or perhaps we may say there can be no tunneling processes
from them.
\vspace{0.3cm}\\
It is instructive to reconsider the FRW tunneling computation in
another coordinate system discussed in previous Sections. Making the
coordinate change  $R:=r a(t)$, the metric assumes the form of
\textit{r-gauge}, namely  \beq ds^2=-(1-H^2R^2)dt^2- 2 H R \,dt dR +
dR^2 + R^2 d\Omega^2 \eeq Note that the metric remains regular on
the horizon and the associated normal metric is of the type
(\ref{nd}). As a result, $\chi=\gamma^{RR}$ and the dynamical
horizon is defined by $H^2R^2_H=1$. Of course, the dynamical surface
gravity has to remain unchanged with respect to (\ref{frw_dyn_sg}).
However, in this particular case, the Kodama vector is very simple,
$K=(1,\overrightarrow 0)$. As a consequence,   the invariant energy
is just $\omega=-\partial_t I$, and the HJ equation for a massless
particle along a radial trajectory reads: \beq \label{hj nolan}
-(\p_t I)^2 - 2 H R (\p_t I)(\p_R I) + (1-H^2R^2)(\p_R I)^2 =0 \,.
\eeq We know already from \textsection II, that the integration
along temporal coordinates gives merely a real contribution.
Thus, an imaginary contribution to the particle action,
comes only from integration along the radial direction. The HJ
equation (\ref{hj nolan}) supplemented of the Kodama energy
constraint, gives \beq (\p_R I)_\pm = -\frac{HR \omega \pm
\sqrt{(HR\omega)^2 + (1-H^2 R^2) \omega^2}}{1-H^2 R^2} \eeq 
Making a null-expansion on the horizon, \beq 0
= \Delta s^2 = -2 \Delta t \Delta R + \Delta R^2 \eeq we see that \beq
(1-H(t)^2 R^2) \approx -2 \left(H + \frac{\dot
H}{2H}\right) \Delta R \equiv 2\kappa_H (R-R_H).\label{step2} \eeq
We finally get the expected results,
 \beq \mbox{Im}\, I_+ = -\int dR
\,\frac{\omega HR\left(1+\sqrt{1+O(R-R_H)}\right)}{-2(H+\dot
  H/2H)(R-R_H-i0)} = - \frac{\pi 
\omega_H}{\kappa_H}\;,\label{im action 2} \eeq with $\kappa_H$
provided by (\ref{step2}).
\vspace{0.3cm}\\
Everything we said above generalizes straightforwardly to models with
non-vanishing spatial 
curvature ($k=0, \pm 1$), except that the surface gravity is given by
the more complicated formula
\beq \kappa_H=-\left( H^2+\frac{1}{2}\dot{H}+\frac{k}{2a^2}\right) R_H
\eeq
where $R_H=(H^2+k/a^2)^{-1/2}$ is the cosmological trapping horizon,
coinciding with the Hubble sphere in the flat case. In
this case, one may note that  $\kappa_H=0$ is possible as soon as
\beq
a(t)=\sqrt{-kt^2+c_1t+c_2}\,,
\eeq
where $c_1$ and $c_2$ are constants. Of course, for $k=+1$, the solution
is real only in a finite range between a big bang and a big crunch. 

\section{Static Black Holes}

Static black hole solutions may be considered as a special case of
dynamical ones. However, they are consistent solutions of
relativistic theory, Einstein or modified alternative theories,  and
they deserve a separate analysis, even though the horizon
tunneling, which we are going to discuss according to  the general
procedure outlined in the previous Sections, is a signal of their
quantum instability  and the static hypothesis we assume is only an
approximation, strictly valid only for limited periods of time.

\subsection{Kodama-Hayward vs Killing Surface Gravity}

Let us consider the Schwarzschild diagonal gauge, which can be
written as \beq ds^2=-V(r)dt^2+\frac{dr^2}{W(r)}+r^2 d\Omega^2\,.
\label{ab} \eeq In this case the function $\chi$ defined in
(\ref{sh}) coincides with $W$. Thus, if we assume that $V$ and $W$
have the same simple zeros, we have, on the horizon defined by
$\chi=W=0$, \beq \frac{V_H}{W_H}= \frac{V'_H}{W'_H}\,, \label{ch}
\eeq via de l'H\^{o}spital ratio rule. These coordinates, as is well
known, are singular on the horizon, so that the null expansion we
have to use is meaningless, the temporal contribution to the action
being ill defined. The use of these singular coordinates  has been
the origin of a large number of papers, containing  several
proposals to deal with the ambiguity \cite{anm}. Recall that in the
original paper by Parikh \& Wilczek on the tunneling method
\cite{parikh}, a clever use of a coordinate system, regular on the
horizon, known as Painlev\'{e}-Gullstrand
  coordinates (PG), was advocated. The general PG gauge reads
\beq ds^2=-V dt^2-2\sqrt{\frac{V}{W}(1-W)}dr dt+dr^2+r^2
d\Omega^2\,. \label{P} \eeq According to our assumption (\ref{ch}),
this gauge is indeed regular on the horizon, being of what we termed
$r$-gauge type. A variant of this gauge is the EF gauge, where an
advanced (retarded) time appears, i.e. \beq ds^2=-V
dt^2-2\sqrt{\frac{V}{W}}dr dt+r^2 d\Omega^2\,. \label{EF} \eeq In
both gauges, we may apply the general analysis and conclude that the
temporal contribution is absent. The Kodama vector is
$K=(\sqrt{\frac{V}{W}},\overrightarrow 0)$ and the related invariant
energy  $\omega= \sqrt{\frac{V}{W}}\partial_t I$. One should note
that the Kodama vector and the invariant energy do not  coincide
with the Killing vector and $\partial_t I$ unless $V=W$. This has
some consequences, since the general theory gives a surface gravity
of \beq \kappa_H=\frac{W'_H}{2}\,, \label{s} \eeq instead of the
surface gravity associated to the Killing vector $\partial_t$ \beq
\kappa_K=\frac{\sqrt{W'_H V'_H}}{2}\,. \eeq More details can be
found in \cite{DiCriscienzo:2008dm}, but for now we just limit
ourselves to note that the tunneling probability  \beq \Gamma
\simeq e^{-2\pi \frac{\omega_H}{\kappa_H}}\,,    \eeq is the
measurable quantity and since \beq
\frac{\omega_H}{\kappa_H}=\frac{E}{\kappa_K}\,, \eeq where
$E=\partial_t I$ is the Killing energy, no contradiction within the
static approximation can be found.  However, as soon as the
space-time becomes dynamic, a Killing vector is useless and the only
invariant energy is $\omega$, indeed.

\subsection{Dilaton-Maxwell-Einstein Black holes }

In this Subsection, we would like to give a brief review
 to black hole static solutions in the form (\ref{ab}), with $W \neq V$. \\
As far as one is concerned with Einstein's GR, solutions of this kind
 are forbidden by well-known uniqueness theorems, cfr.\ \cite{FN}.
However, it is not difficult to face them in alternative theories of
 gravity, a label which actually contains a lot of different material.
In order to be more precise, let us consider the specific example
 provided by the so-called Dilaton-Maxwell-Gravity
(DMG) as in \cite{mann} and \cite{gibbons,garf,horo,Chan:1995}. We typically
start by an action such as \cite{Chan:1995}, \beq I = \int d^4 x \sqrt{-g}
\left(R - 2 (\nabla\phi)^2 - V(\phi) - e^{-2
 \xi \phi} F^2\right)\label{daction}
\eeq
where $R$ is the scalar curvature, $F^2= F_{ab}F^{ab}$ and $\xi$
 governs the coupling of the dilaton with the Maxwell field.
Varying the action (\ref{daction}) with respect to the metric, Maxwell
 and dilaton fields yields the EOM for the respective fields.
It is easier if we consider Maxwell fields generated by an isolated
 electric charge. Then, the ansatz
\beq
ds^2 = - U(r)dt^2 + U^{-1}(r) dr^2 + H^2(r)d\Omega^2 \label{dmetric}
\eeq
for the line element, satisfies the EOM for the metric under very
 general conditions.\footnote{The Maxwell field then reads
$F_{t r} = \frac{Qe^{2\xi\phi}}{H^2(r)}$, $Q$ being the electric charge.}

Assuming $b < a$, a first class of solutions is given by \cite{garf}
\beq
U(r)=\frac{1-\frac{a}{r}}{1-\frac{b}{r}}\, \qquad \&\qquad H(r)=r^2-br\,,
\eeq
another one is instead
\beq
U(r)= 1-\frac{a}{r}\,,\qquad \&\qquad H(r)=r^2-br\,.
\eeq
For a quite general class of solutions, but without any attempt of
universality, the $U(r)$ function in (\ref{dmetric}) can be
expressed in terms of $H(r)$ and the dilaton as
\beq
U(r) =\frac{\xi r + \eta}{H^2(r) \phi'(r) + \xi H(r) H'(r)} \label{U}.
\eeq
where $\eta$ is an integration constant and the prime denotes
derivation with respect to the argument.\\
Next, consider a conformal transformation
\beq
d\tilde s^2 = \Omega^2 ds^2
\eeq
so that
\beq
\Omega^2(r) = \frac{r^2}{H^2(r)}. \label{ct}
\eeq
The conformal metric looks like (\ref{ab}),
\beq
d\tilde s^2 = - V(r) dt^2 + W^{-1}(r) dr^2 + r^2d\omega_2^2
\eeq
with
\beq
V(r) &=& \frac{r^2 U(r)}{H^2(r)} = \frac{r^2(\xi r +\eta)}{H^4 \omega
  + \xi H^3 H'} \label{appV}\\
W(r) &=& \frac{U(r) H^2(r)}{r^2} = \frac{H(\xi r+\eta)}{r^2(H\omega +
  \xi H')} \label{appW}.
\eeq If the spacetime possesses a horizon, this will be located at
$r_0$, s.t.\  $W(r_0) =0$, i.e.\ where
\begin{enumerate}
\item $U(r_0) =0 $, that is $r_0 = - \eta/\xi >0$;
\item $H(r_0) =0$. Note that in this case
\beq
\lim_{r\rightarrow r_0} V(r) = \infty \qquad \lim_{r\rightarrow r_0}
\Omega^2(r) = \infty\,,
\eeq
that is, the conformal transformation becomes singular.
\end{enumerate}
It is simply a question of algebra to compute the Kodama-Hayward
surface gravity and see that it is well-defined in both cases 1 and
2. But, with regard to the Killing surface gravity, we have to
distinguish carefully between the two cases under examination.
Indeed, case 1 is easy to treat and gives $\kappa_K = U'_0/2$:
something we had to expect from the very beginning in consideration
of the well-established result according to which the Killing
Hawking temperature $\Theta=\kappa_K/2\pi$ is conformally invariant
\cite{Jacobson:1993pf}. But things go radically different in case 2
of a singular conformal transformation, where the same definition of
 a Killing surface gravity, positive and finite on the horizon becomes,
 at least in the general case, questionable.

This analysis, however, is sufficient to shed new light on the stringy
black hole puzzle.
Start in fact from the metric (\ref{dmetric}) with, for example,
 $U(r) = 1-a/r$, $a=const$ and $H(r) = \sqrt{r(r-b)}$, $b<a$.
 Perform a conformal transformation in order to get the GHS solution
 \cite{garf}.
 Since $b<a$, $r_0=a$ is still an event horizon. The Killing surface
 gravity,  being invariant under conformal transformation,
 does not feel the new physics introduced by the conformal
 transformation. Thus no change has to be expected in the extremal
 limit.
The story is different for Hayward's surface gravity that goes like $\kappa_H
\propto H_0^2$  and vanishes whenever the conformal factor vanishes, e.g. in
the extremal limit for GHS solution.

\subsection{The Lema\^{\i}tre-Rylov Gauge} 

As further example, let us consider the  Schwarzschild spacetime in coordinates
$(t, r, \theta, \varphi)$ such that the line element can be expressed
as 
\begin{equation}
 ds^2 =-dt^2 + \frac{dr^2}{B} + (r_g B)^2 d\Omega^2\;, \label{lemaitre}
\end{equation}
where $r_g = 2m$ is the usual gravitational radius, and
\begin{equation}
 B(t,r) := \left[\frac{3}{2r_g}(r-t)\right]^{\frac{2}{3}}\;.
\end{equation}
We shall refer to these coordinates as the Lema\^{\i}tre-Rylov
  gauge. This is indeed an interesting (time-dependent) gauge since -
contrary for example to isotropic coordinates - $(t,r)$ extend beyond
the gravitational radius, $r < r_g$. It is an explicit example of
synchronous gauge.\\  
Notice further that \cite{FN}:
\begin{enumerate}
 \item The spacetime singularity is located at $r= t$;
 \item The horizon is located in correspondence of 
\begin{equation}
 B(t,r)\big\vert_H = 1\,,\,\,\quad (r_H-t)=\frac{2 r_g}{3}\,.
\end{equation}
\item Outgoing particles are such that $\frac{dt}{dr} <0$; ingoing
  particles instead have $\frac{dt}{dr} >0$ (cfr. spacetime diagram in
  \cite{FN} for example).  
\end{enumerate}
A detailed calculation can now be provided.\\
According to the general procedure outlined above, 
\begin{equation}
  \gamma_{ij} = \left(\begin{array}{cc}
-1 & 0 \\
0 & B^{-1}
\end{array}\right)\;,\qquad \sqrt{-\gamma} = B^{-1/2}\;,
\end{equation}
and
\begin{equation}
  \gamma^{ij} = \left(\begin{array}{cc}
-1 & 0 \\
0 & B
\end{array}\right)\;.
\end{equation}
Making use of (\ref{H}), a direct computation leads to
$k_H=\frac{1}{2r_g}$, as expected.  
The explicit form of the Kodama vector $K$ in terms of the
Lema\^{\i}tre gauge is 
\begin{equation}
 K = (1,1,0,0)\;,
\end{equation}
so that the particle's energy is
\begin{equation}
 \omega = -\partial_t I - \partial_r I\;.
\end{equation}
The H-J equation for radially moving particles is
\begin{equation}
 -(\partial_t I)^2 + B (\partial_r I)^2 =0\,,
\end{equation}
and for the outgoing particles, we take
\begin{equation}
 \partial_r I = - \frac{\partial_t I}{\sqrt{B}}\,.
\end{equation}
Thus
\beq
\p_r I=\frac{\omega (\sqrt{B}+1)}{B-1}\,.
\eeq
The null expantion condition gives for the outgoing particle $\Delta
t=-\frac{\Delta r}{\sqrt{B}}$, namely 
we have 
\begin{equation}
I= 2 \int_{\gamma} dr \, (\partial_r I)_+ = 2 \int_{\gamma} d r \omega \,
 \frac{\sqrt{B}+1}{B-1}.\label{Lem3} 
\end{equation}
Let us define $\mathscr A(t,r) := B-1$ and expand it along the
null, radial, outgoing geodesic close to the horizon: 
\beq
B-1\simeq \p_r B_H \Delta r+\p_t B_H \Delta t= 2 \p_r B_H \Delta
r=\frac{2}{r_g}(r-r_H)\,. 
\eeq
As a result
\beq
I=r_g \int_{\gamma} \frac{\omega  (\sqrt{B}-1)}{(r-r_H-i0)}  \quad
\Longrightarrow \quad 
 Im\, I = \frac{\pi \omega}{k_H}=4\pi m \omega \label{ok} 
\eeq
which provides the correct Hawking temperature from the hole.
The Lema\^{\i}tre-Rylov gauge can be generalized beyond Schwarzschild
spacetime. In general the metric we shall deal with is 
\beq
ds^2=- dt^2+(1-V(R))dr^2+R^2 d\Omega^2\,, 
\label{LR}
\eeq
where $V(R)$ and the areal radius $R$ are function of $r \pm t$, via
the inversion of the relation 
\beq
r \pm t=G(R)= \mp \int dR \sqrt{\frac{1}{WV}} \left(
\sqrt{\frac{1}{1-V}}-  \sqrt{1-V} \right) \,.  
\eeq
We are dealing with a dynamical synchronous gauge and we may apply the
general formalism seen above.  A straightforward calculation  
leads again to  (\ref{s}) \footnote{As an exercise, the interested
  reader could be curious to see what happens by introducing  
Kruskal-Szekeres generalized coordinates for the metric
(\ref{ab}). The computation is rather long and tedius, but still
confirms,  as it must be, the general covariance of Hayward's surface
gravity.}.

\section{Conclusions}

In the last few years, many different proposals have been suggested in
 order to give a universal prescription for the Hawking temperature of
 certain dynamical spacetimes, especially those endowed with future
 trapping horizons. In such cases, for
 example, one lacks the Kubo-Martin-Schwinger (KMS) condition so
 successful for equilibrium
 states. Nor is there generally available a analytic continuation to
 Euclidean signature within which one can judge the periodicity of
 Euclidean time, essentially equivalent to the KMS condition. In
 fact, one can even doubt that a temperature with the usual meaning is
 generally possible or useful at all. The final result is that time by
 time many prescriptions were
 proposed which either were not applicable to all the desired cases
 one can have in mind, or simply were invented ad hoc to keep the
 peace  with the most well known cases. For example, for a slowly
 varying Schwarzschild black hole it seemed natural to keep the
 temperature equal to the instantaneous value $1/8\pi M(t)$
 characterizing the static black hole, although this is not so obvious.

On the other hand we are all accustomed to the remarkably universal
 properties exhibited by black holes so, moved by the wish of
 extending some of these to a dynamical regime, we have
 written this note in order to clarify (hopefully) the  so-called HJ
 method for horizon tunneling.

The playground has been given by the class of spherically symmetric
 spacetimes, either static or dynamical, and within this class
we can draw the conclusion that in order for the HJ method to work
 properly: (i) regular coordinates at the horizon are necessary;
(ii) an appropriate notion of particle energy, that it should be a
 scalar, can be  implemented at the level of HJ equation in such a
 way that imaginary contributions to particle action of outgoing
 particles arise after a near-horizon approximation; (iii) we have
shown the relevance of the null-expansion of (to all purposes,
 massless) particles through the horizon in order to reconstruct fully
the particle action. In turn, it is the particle action responsible for
 the particle production rate (\ref{prob}).

As a  result, it should also be clear that a contribution to the
 rate from the temporal part of the integration, that is from the
 piece $\int_{\gamma} \partial_t I dt$ of the basic tunneling rate, is
 generally present but gauge dependent; it is only the full
 integration along $\overrightarrow\gamma$ that produces a gauge
 invariant result. So for 
 example, in certain gauges there is a temporal contribution, in some
 other gauge there is not.

Finally, in static spacetimes violating the weak energy condition it
seems that the
 choice between Killing and Kodama-Hayward cannot be decided at the
 formal level, although it may be remarked that the difference amounts
only to a trivial scaling, albeit one which can have less trivial
 effects on extremal black holes. It may also be noted that, as soon as a
 static black hole starts evaporating, the spacetime
 ceases to be rigorously static and only the dynamical picture
survives, a picture where the Killing vector with its associated
 energy/surface gravity/temperature plays a much more minor role.
\medskip \\
RDC wishes to thank the \textquotedblleft Center for Astrophysics -
Shanghai Normal University'' for warm hospitality. SAH was supported
by the National Natural Science Foundation of China under grants
10375081, 10473007 and 10771140, by Shanghai Municipal Education
Commission under grant 06DZ111, and by Shanghai Normal University
under grant PL609.

\end{document}